# A Mobile Device Prototype Application for the Detection and Prediction of Node Faults in Wireless Sensor Networks

Anthony Marcus, Ionut Cardei, Borko Furht
Department of Computer and Electrical Engineering and Computer Science
Florida Atlantic University
Boca Raton, Florida U.S.A.
{ amarcu10, icardei, bfurht }@fau.edu

Osman Salem and Ahmed Mehaoua
UFR de Mathématiques et Informatique
Université Paris Descartes -LIPADE Paris, France
POSTECH, Division of IT Convergence Engineering, Korea
{ osman.salem, ahmed.mehaoua }@parisdescartes.fr

*Abstract* — **Various implementations of wireless sensor networks (i.e. personal area-, wireless body area- networks) are prone to node and network failures by such characteristics as limited node energy resources and hardware damage incurred from their surrounding environment (i.e. flooding, forest fires, a patient falling). This may jeopardize their reliability to act as early warning systems, monitoring systems for patients and athletes, and industrial and environmental observation networks. Following the current trend and widespread use of hand held, mobile communication devices, we outline an application architecture designed to detect and predict faulty nodes in wireless sensor networks. Furthermore, we implement our design as a proof of concept prototype for Android-based smartphones, which may be extended to develop other applications used for monitoring networked wireless personal area and body sensors used in other capacities. We have conducted several preliminary experiments to demonstrate the use of our design, which is capable of monitoring networks of wireless sensor devices and predicting node faults based on several localized metrics. As attributes of such networks may change over time, any models generated when the application is initialized must be updated periodically such that the applied machine learning algorithm maintains high levels of both accuracy and precision. The application is designed to discover node faults and, once identified, alert the user so that appropriate action may be taken.**

## I. INTRODUCTION

Our desire to gather and disseminate information, regardless of location, has motivated researchers and engineers to develop novel hardware technologies which allow communication over vast distance, from any location through satellite links and terrestrial, high throughput networks. These new, super powerful, handheld smart devices have evolved to the point where those available today exceed the computational power found in the average desktop computer. With such power at our fingertips, computer scientists and engineers are able to create software applications with high complexity that are capable of almost anything imaginable from video conferencing and remote area observation and monitoring to voice recognition and artificial intelligence. Persistent research into new possible frameworks and architectures for these applications continues to generate novel concepts and systems.

Scientific research as to how various species learn and gain knowledge has resulted in learning profiles derived from these studies. Researchers have studied these volumes of information in great depth to increase understanding and attempt to extract the underlying ideologies and algorithms involved. It is these resulting methods which are then integrated into machine learning applications and software solutions. When data mining and machine learning (DMML) is applied in areas involving recognition, enhancement, diagnosis, planning, robot control, prediction, and the like, the algorithms involved contribute to the overall, "artificial intelligence (AI)" of the system.

Recent advancements in low cost, micro processing and sensing technologies have generated enormous interest and allowed the widespread use of networks of computational sensor units capable of wireless communication using various protocols over great distances. Following the explosive trend of wireless sensor network (WSN) applications and implementations, many developers look to increase the reliability and quality of service (QoS) required by these systems using a variety of techniques and novel ideas. In [1] a general architecture for a WSN is outlined, which may describe a small network with few nodes and a single sink such as in personal area networks (PAN) [2, 3, 4] and wireless body area networks (WBAN) [5, 6, 7, 8] or may have numerous nodes arranged in hundreds of clusters used to monitor large areas [9, 10].

WSNs play a critical role in monitoring hazardous environments where it is unsafe to have humans permanently stationed nearby. Researchers and scientists traversing rough, dangerous terrain in order to gain access to a WSNs location and gather and evaluate data from and about the network, need not be excessively encumbered with heavy gear and supplies. Our objective is to create a smartphone based utility which is capable of mining network metrics and issuing alerts about actual and trends toward faults. This would ensure maximum mobility, by minimizing encumbrance associated with necessary equipment, without having to compromise the performance or accuracy of the WSN analysis.

Using our application architecture we developed and implemented a proof of concept prototype, sensor management and regression tool (SMART), for node fault detection and prediction in an environmental WSN. Individual wireless sensor nodes are used to measure specific characteristics of an object or environment. These measurements are gathered and stored on the smartphone from each node in the network in real time after successfully creating a TCP connection with a WSN base station. As data is retrieved and stored, the system applies DMML algorithms to extract important attributes and correlations from the data. These are then used to build learning models for each sensor and each node. The models are then used for node classification and the prediction of values for each sensor. SMART then looks for faulty classified nodes

This work was supported in part by NSF grants CCF-0545488 and OISE-0730065 with additional support from the Korean Science and Engineering Foundation, under the World Class University (WCU) program.

and large deviations between predicted and actual values. If the later value is greater than the specified percentage threshold and all other values lie within the threshold, we mark the outlier sensor metric as faulty. The application then issues an alert to the user, including the node identifier for those which are classified as faulty and when specific sensors breach the percentage threshold.

Our initial experiments were utilized to determine which numeric prediction algorithm was best suited for our application as there are several constraints associated with mobile platforms including the relation between time, energy, and complexity. In other words, the greater the complexity, the more time and energy is required to complete the task. Therefore, the primary focus of our early experimentation was to determine which algorithm gave us the best balance while maintaining highly accurate results. Further experimentation is planned, to test for possible unknown limitations of our architecture.

This paper looks first at some related work in Section II, followed by a description of the prototype design in Section III and experiment in IV. In Section V we describe our observations and data, followed by conclusions in Section VI and our plans for future work in VII.

## II. RELATED WORK

Since the wide spread implementation and use of WSNs, researchers continue to develop new methods to refine the capabilities of these networks. There have been numerous projects which attempt to reduce and detect network faults and hardware malfunction and degradation, insuring that these networks continue to operate optimally. WSN faults may occur for several reasons such as energy depletion, component fragility, broken links, partitioning, and more. Recently several projects have been conducted which are primarily focused on detecting WSN faults, using a variety of novel methods, some of which gave us inspiration to create our application.

In [11] the authors describe an algorithm which is capable of detecting node faults in WSNs. The approach, named FIND, ranks each node based on the sensed readings and their physical distances after each locally detected event. Distances are defined and measured according to the signal attenuation, as distance has a direct effect on this parameter, assuming that sensor readings somewhat reflect the corresponding distance. A node is labeled as faulty when there is a significant mismatch between the signal and data rankings. The proposed theoretical model shows that the average distance ranking is a provable indicator of possible data faults.

M. Lee et al. [12] developed a fault model capable of identifying and detecting node faults, under the assumption that each node has a similar transmission range and equivalent hardware functionality. The algorithm is based on comparisons between neighboring nodes and broadcast decisions made by individual nodes. Time redundancy is also considered in order to tolerate any transient faults found both in sensing and communication. To minimize any delays resulting from the time redundancy, a sliding window is employed which stores the previous results.

Other researchers focus on the development of novel methods and concepts, often revealing new application areas in the field of DMML. One such project [13] exploits DMML algorithms to determine WSN reliability and faulty hardware in the network. They present a novel approach which applies Logistic Regression [14] to determine the reliability of a WSN. Placing emphasis on the concept of reliability and its importance for WSNs, their method is applied to classify individual nodes as faulty or not faulty, more specifically, they consider reliability based on a specific type of network lifetime.

Another novel application described in [15], implemented the k-NN algorithm to estimate non-linear relationships between multipath signal parameters and a mobile terminal position. To validate their solution, simulations were conducted using a realistic multipath scattering environment. These systems have several areas where they no longer function optimally, such as time synchronization problems and non-line of sight and multipath errors [16, 17].

Producers of mobile products in recent years have begun to integrate MIMO communication technologies into their products [18]. This particular application exploits the characteristics associated with MIMO such that it may, using a single base station, determine the geographical location of a mobile target. Applying advanced array signal processing [19, 20], they estimate the angle of arrival, angle of departure, and delay of arrival of MIMO signals. To train the classifier, they collected measurements at various locations, each location an instance having n measurement attributes. A mapping model was then generated from this data applying weighted k-NN and Euclidean distance.

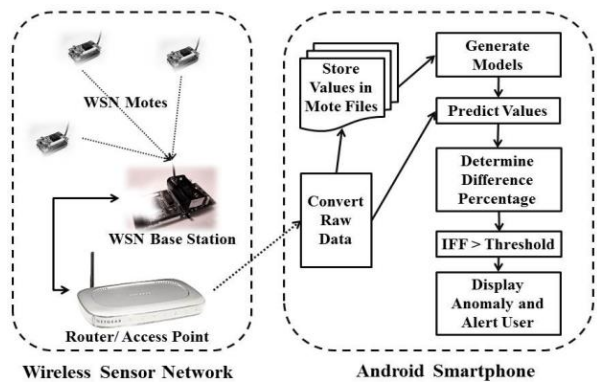

*Figure 1:* **Application architecture showing connected WSN components and generalized internal working of Android SMART software.**

Each of these projects incorporate specialized algorithms to detect and find attributes and correlations associated with WSNs which may have a detrimental impact on network functionality and reliability. After observing the various techniques these projects utilize to achieve their objective, we integrate similar concepts into our application design. While extensive research into fault detection in WSNs has been conducted, each project requires a laptop or desktop computer. Our design is such that we may conduct the analysis onboard a smartphone which gives us the capability to find network faults locally, requiring no internet connection, and increasing mobility by eliminating the need carry cumbersome computational equipment to remote WSN locations.

## III. PROTOTYPE DESIGN

Our application architecture, shown in Figure 1, implements four main functions; data acquisition, data consumption, model generation, and real time classification and prediction. In data acquisition, WSN nodes monitor objects or a physical environment. The object or environment to be monitored may include living or mechanical entities, indoor/ outdoor systems or environments, or combinations of each. WSN nodes read and transmit various metrics from their sensed surroundings to a base station that forwards the data to another system. Data consumption uses the interface to process this raw data into usable engineering units and then stores it locally. A monitoring application identifies interesting conditions and events, alerting the user if necessary. In model generation, SMART uses DMML algorithms which mine and analyze the stored metrics and generate prediction models for forecasting node health. Each sensed attribute, in each incoming instance, uses its prediction model to conduct real time numeric prediction. For the data consumption, model generation, and real time classification and prediction, the smartphone is running our custom built application.

SMART reads, stores, and mines WSN sensor node metrics from each node in the network. It then uses saved data to build prediction models for each sensor using methods from the Weka API [23]. After the initial models have been generated, it uses them to predict and identify probable weaknesses and node failures in the network. The concept is such that in a WSN, prior to the occurrence of failures, warnings can be generated and proper action taken such as hardware replacement or re-routing of network data. We use a short term window to observe the spatial aspect of the WSN for rapid prediction and location of faulty network nodes and to ensure that these models and metric thresholds are updated periodically such that the predictive advantages of the application are preserved. A long term, history based window is employed for the temporal analysis of the WSN, which enables the confirmation of node faults found in the short window and also analysis of trends in the network nodes over time. Our design is flexible such that both PANs and WBANs, used in areas such as healthcare and patient monitoring and sport activities, and WSNs, used for environmental monitoring, may be analyzed locally in real time for reliability and failures.

Our application is designed for the Android platform as this operating system allows developers greater freedom to access specific components and simplifies access and integration with other required functionality from various Java libraries. Our WSN consists of Crossbow sensor components as this system was readily available for implementation. Sensor motes are programmed in the nesC language for the TinyOS runtime environment designed specifically for low frequency computer components. To test our proof of concept, we conducted some preliminary experiments using a variety of DMML numeric prediction algorithms to find the one best suited for our particular WSN scenario.

## IV. EXPERIMENTAL DESIGN

To demonstrate the use of SMART for a WSN monitoring the environment, we developed an Android [24] application and custom firmware for the WSN motes. With our firmware installed, the WSN nodes report specific sensed metrics of their immediate environment to the base station. The transmitted attributes include light, temperature, acceleration in a 2D plane, and voltage of the power supply. These are then packaged in AM message format and transmitted to the base station. The base station receives and buffers these packets for access by SMART. Our experimental WSN utilized a small network of motes to provide us with proof of concept, but many environmental applications consist of numerous nodes. The scalability of the application is limited only by the computational ability of the smart device used for the DMML operations. We also write the incoming WSN data to files rather than storing everything in the RAM memory to reserve it for the DMML segment of the program.

### a. Materials

*WSN Hardware:*

- Crossbow MICAz MPR2400CA motes [25, 26] equipped with a 2.4 GHz, IEEE 802.15.4 compliant ZigBee transceiver, and 4Kbytes of RAM
- Crossbow MTS310 [27] sensor boards equipped with light, temperature, 2D accelerometer, 2D magnetic flux sensor, microphone and sounder
- Crossbow MIB600CA [28] with Ethernet (10/100 Base-T) serial connectivity to bridge the "wired" and "wireless" segments of the network
- Netgear Wi-Fi 802.11 b/g [29] access point

*SMART Interface Hardware:*

- LG Optimus 2X (P990) [30] equipped with a 1GHz, dual core Tegra processor, Wi-Fi 802.11 b/g/n network card, 512MB RAM, 40GB storage, running Android 2.3.4.

### b. Preparation

To prepare each component for use in the experiment, we describe several mandatory requirements:

*With respect to the smartphone:*

- Needs to be running the Android operating system version 2.3.4 or later
- Needs to have its wireless capability enabled prior to running the application
- Needs to be successfully connected to the access point of the WSN base station it will monitor
- Needs to have the option 'install third party applications' enabled in the phones settings
- Needs to have ample storage (>5Gb in our case) for the .apk and data files of the application which is dependent on the number of nodes in the network

*With respect to the WSN:*

- Nodes used for sensing need to have the appropriate firmware installed
- Nodes must have batteries installed with adequate power

- The base station mote must have the appropriate firmware installed
- The base station needs to be connected to the MIB600 Ethernet gateway
- The gateway needs to be connected via Ethernet to a router or access point

c. **System Logic**

The logic flow of our application is shown in Figure 2, where the smartphone first creates a TCP connection and requests permission to access the WSN base station. Once access is granted, it begins to intercept the raw message packets from the network every second. Each time a new message is received SMART removes the message wrapper and parses the raw contents. It then converts the raw data into useful engineering units for each individual sensor using specific formulas for each attribute. After conversion, each metric is saved as an attribute in each instance for each node in the network. Each node has its own short term file which consists of instance readings every second for one hour, a total of 3600 instances. Every minute, the average value for the previous 60 readings is stored in a long term file which consists of each averaged instance for an entire 24 hour period or 1440 instances. We use the long window to incorporate daily sensor trends into our analysis, for example the light and temperature sensors will fluctuate with the time of day as the Sun's movement will have an impact on the reading.

Each time a new instance is added to a nodes' short and long files, we include a timestamp so that we may examine correlations between the time of day and particular sensor values. At the end of each node's short time period, we use these 3600 readings to generate new prediction models for each sensor measured in the analysis. Updating these models every hour allows us to incorporate any spatial changes to sensor readings, increasing the accuracy of our application.

If we want to discern between sensor changes occurring, for example, at sunset from those that occur in a forest fire, we must observe and extract network trends over a longer period of time. The purpose of observing the long time period is such that we may incorporate periodic sensor trends which would not be identified in the short window. We generate the long window values by taking the averages for each sensor using the values sensed from the previous minute (60 readings). Again each node has an associated long window file, which encompasses the past 24 hour period. At the end of each period, models are generated or rebuilt for each sensor, which are then used for long window numeric prediction and reflect the long term trends for each sensed attribute of each node in the network.

After the initial short window models have been generated once the first hour of data has been gathered, incoming instances are fed through these models and for each sensed attribute a predicted value is calculated which we utilize to determine faults in the WSN. Performance and reliability analysis of DMML algorithms is always dependent on the type of data analyzed and enforces the need for testing in order to determine the best choice for a particular application. For our initial experiments, we tested a variety of DMML algorithms including Linear Regression [31], Decision Stump [32], Decision Table [33], k-nearest neighbor (IBk, k-NN) [34], and M5P [35].

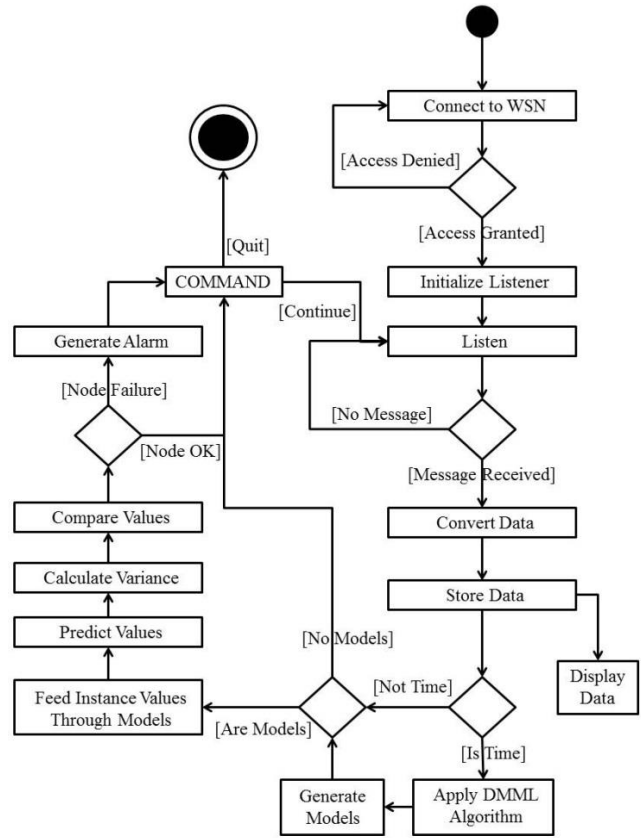

*Figure 2:* **Activity diagram showing the SMART application logic.**

Once the predicted values have been generated, they are compared to the actual values to determine the difference percentage. This measurement is then compared to the percentage of variance thresholds we allow for each particular sensor, for example, the threshold of variance for the temperature sensor is 5%. If, in this particular case, the difference between the predicted and actual value is greater than 5% and all the other sensed attribute values in this instance are within their specified threshold boundaries, we feed the instance through the long window models if they exist. Percentages associated with the threshold boundaries, depend on the WSN implementation (i.e. environment, observed object). If no models have been generated for the long term window, then the sensor is marked immediately as faulty and an alert is triggered. If there are long term models, the instance is fed through the models to observe if the difference still exists. If this anomaly still exists, an alert is generated and the user notified. The alert generated displays to the user the mote number and the predicted faulty sensor.

At the end of every hour, the short term models are updated and the previous instance and model files are overwritten. Similarly at the end of every long cycle the models are updated and files overwritten. These temporal parameters for both short and long term files have been set such that the file size limitations of the Android platform are not exceeded. We also use these windows to assist in the identification of short and long term trends in the network, as the use of only one of these windows would reduce the accuracy of the application.

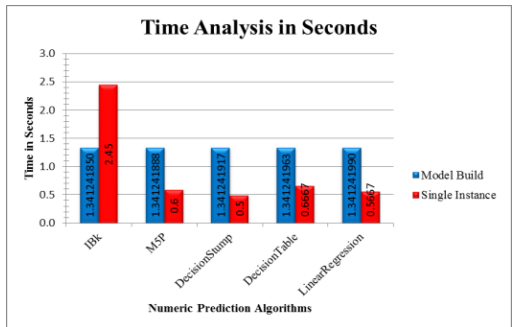

*Figure 3: CPU* **time requirements for both model generation and single instance numeric prediction using the generated models, for each tested algorithm.**

## V. OBSERVATIONS AND DATA

In this section we outline our observations when conducting preliminary experiments with our proof of concept application. Our first tests were conducted to determine which prediction algorithm was able to generate sensor models and predict attribute values in the shortest time period. The results, shown in Figure 3, reveal that Decision Stump had a negligible difference to generate the initial models from the training data and the shortest interval needed to conduct real time prediction. These times would be directly affected by the number of instances in the training data, and would also vary according to the smartphones' capabilities (i.e. CPU, RAM).

WSN message packets are received from the motes in succession. Each node's information is displayed on the smartphone and each instance stored in a separate file for each node. Functionality from the Weka libraries properly generates prediction models for each attribute. When a short term cycle of data has been analyzed and modeled, real time readings received from the nodes are fed directly through the appropriate model for classification. If the variance threshold is exceeded for any attribute's predicted value when fed through its associated short and long term model, an alarm is generated to notify the user of possible node failure.

Further tests were conducted so that we may observe the absolute error distribution, shown in Figure 4, for each algorithm. This graph shows that the distribution for Decision Stump had the lowest absolute error above 30%, and had many predicted values within 10% of the actual value. Considering that, for smartphone implementations, we need an algorithm which maintains a low error rate and energy consumption, while having a quick analysis time, Decision Stump demonstrated the best overall performance. This is clearly shown in Table 1, where we observe the energy consumption associated with each instance. We approximate the total energy consumption per instance based on the percentages found in the smartphone settings which show battery consumption for display and CPU usage in percent and time used, combined with Wi-Fi's use of 330mAh reported by the Android API and a transfer rate of approximately 1Mbps, knowing each message size is 224 bits. As these were preliminary tests with our application, having this additional knowledge allows us to update our design for future testing and increased accuracy. Again, these results show Decision Stump to be the best performing algorithm.

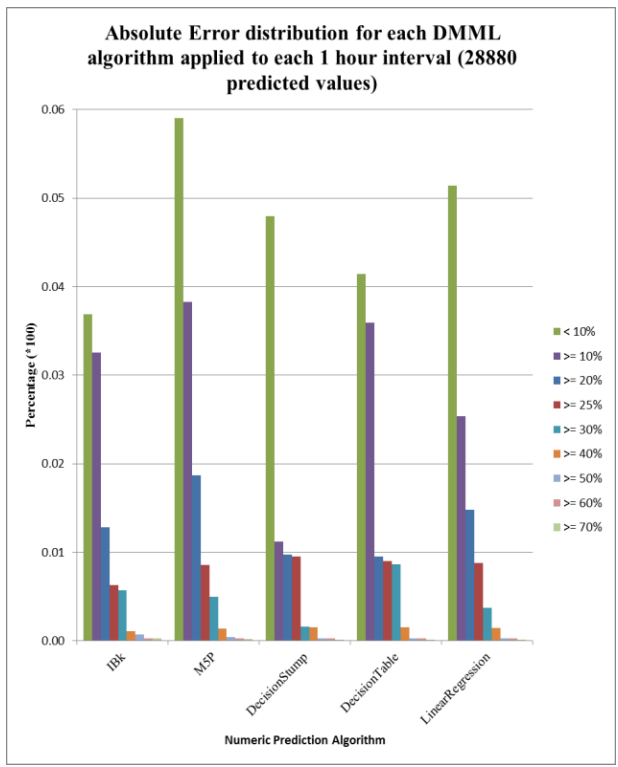

*Figure 4:* **Absolute error distribution for each tested numeric prediction algorithm.**

|  | Instance Prediction Time in Seconds | Approximate Energy Consumption per Instance in Joules | Total Error Rate Percentage |
|---|---|---|---|
| IBk | 2.45 | 0.03988008 | 5.98338 |
| M5P | 0.6 | 0.00976356 | 7.278393 |
| DecisionStump | **0.5** | **0.00813852** | **3.43144** |
| DecisionTable | 0.6667 | 0.0108558 | 6.533934 |
| LinearRegression | 0.5667 | 0.00923076 | 5.481302 |

*Table 1:* **Experimental results for prediction time, approximate energy consumption, and error percentages for each algorithm with the overall best performance in bold.**

## VI. CONCLUSIONS

Breakthroughs in mobile hardware architecture are constantly marketed which allow greater local computation in a lightweight, hand held package. As a result, developers are creating applications which have highly complex architectures and capabilities and push the thresholds of current DMML research and technology. We described a variety of DMML applications which greatly improve both accuracy and functionality of systems in many scenarios using a variety of complex formulas and innovative statistical techniques.

As wireless and sensor technologies are following a similar trend, engineers and researchers continually develop new technologies and application areas for WSNs in everything from individualized athlete and patient monitoring to hazardous environment observation. Sensor node anomaly and failure detection is extremely important to WSNs where reliability and QoS assurance is critical.

Our proof of concept application is designed to operate on small, portable smart devices, detect node faults in a WSN, and alert the user as to any anomalous predictions. Early experimental results have demonstrated the ability of the application to pinpoint node hardware faults in a WSN. Further experiments and refinement of the software are planned such that we may increase the accuracy, while reducing the time and energy needed for analysis.

## VII. Future Work

To further test our prototype design we are currently developing prototype systems for two other WSN application scenarios. As the reliability of metrics gathered by PANs used in the medical industry for patient observation and long term monitoring is extremely critical, we apply SMART to our prototype design which is capable of analyzing sensed patient attributes in real time.

Secondly, we are prototyping and testing an application which will predict broken communication and control the flow of data between an ocean-based platform and a land-based component using various sensor components. Maintaining high levels of both reliability and QoS for wireless communication between components in the OCTT [36] architecture is critical to individuals testing prototype turbines for the purposes of harvesting renewable energy from oceanic currents.